# Dynamic Metrics of Heart Rate Variability

Clinton F. Goss[†] and Eric B. Miller[‡]

[†]Westport, CT, USA. Email: clint@goss.com
[‡]Department of Music Therapy, John J. Cali School of Music, Montclair State University, Montclair, NJ 07043; Email: miller@biofeedback.net

**ABSTRACT**

Numerous metrics of heart rate variability (HRV) have been described, analyzed, and compared in the literature. However, they rarely cover the actual metrics used in a class of HRV data acquisition devices – those designed primarily to produce real-time metrics. This paper characterizes a class of metrics that we term *dynamic metrics*. We also report the results of a pilot study which compares one such dynamic metric, based on photoplethysmographic data using a moving sampling window set to the length of an estimated breath cycle (EBC), with established HRV metrics. The results show high correlation coefficients between the dynamic EBC metrics and the established static SDNN metric (standard deviation of Normal-to-Normal) based on electrocardiography. These results demonstrate the usefulness of data acquisition devices designed for real-time metrics.

## Introduction

Metrics of heart rate variability (HRV) are widely used in medical diagnostics, physiological research, and biofeedback training (TaskForce, 1996; Wheat & Larkin, 2010). HRV metrics are typically based on non-invasive measurements of cardiac function using electrocardiography (ECG) or photoplethysmography (PPG). A specific point is identified in each cardiac cycle, and the inter-beat interval (IBI) between those points in adjacent ECG heartbeats or PPG pulsebeats is measured.

Time intervals derived from ECG provide the input to a function that produces the HRV metric. Functions defined over PPG time intervals are often termed metrics of pulse rate variability (PRV).

A broad range of linear and non-linear HRV and PRV metrics have been developed based on statistical, spectral, geometric, neural network, and vector analysis techniques.[1] These metrics offer various tradeoffs in their computational complexity, applicability to real-time analysis, tolerance for physiological artifact and measurement artifact, and significance with respect to particular medical or physiological conditions.

Most of HRV and PRV metrics described in published literature operate directly on a set of IBIs taken over a given length of time – we use the term *static metrics* in this article for these metrics. However, a wide selection of devices are available that provide metrics constructed in fundamentally different ways. These devices are typically designed for real-time reporting of metrics. Hence, we use the term *dynamic metrics* for the class of metrics provided by real-time data acquisition devices.

This article characterizes dynamic metrics and reports the results of a pilot study which compares one such dynamic metric with established HRV metrics.

### ECG and PPG Metrics

In this article, we differentiate ECG-based metrics from PPG-based metrics using *e* and *p* postscripts, respectively. For example: SDNN*e* and SDNN*p* are variants of the widely-used SDNN metric of HRV. SDNN*e* is the standard deviation of the intervals between adjacent R peaks of the QRS complex derived from normalized ECG measurements recorded over a given time period. SDNN*p* is the corresponding metric based on peaks in normalized PPG measurements.

---

[1] An informal survey of published literature conducted by the second author has cataloged 69 distinct metrics, without distinguishing between ECG and PPG variants.

Because of the relative ease of collecting PPG measurements in many situations, PRV metrics are often preferred to ECG-based HRV metrics. A number of studies have examined the relationship between the ECG and PPG variants of specific metrics, in particular SDNN (Johnston & Mendelson, 2005; Selvaraj, Jaryal, Santhosh, Deepak, & Anand, 2008; Teng & Zhang, 2003).

### Dynamic Metrics

Dynamic metrics introduce a number of characteristics not typically found in static metrics:

**Span**. This extends the definition of IBI beyond the limitation of using adjacent cardiac cycles. The metric may be based on an IBI where the interval is a *span* measured between a given number, C, of cardiac cycles apart (span = C cycles) or between the cardiac cycles that are at least a given number, S, of seconds apart (span = S seconds).

An example is a metric defined over a heart rate (HR) derived from an IBI that spans 10 cardiac cycles.

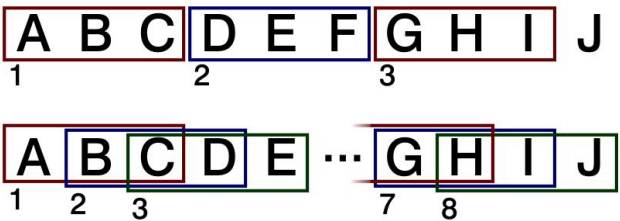

**Figure 1**. Two examples of sampling window progressions using sampling windows of size *W* = 3. A through J are inter-beat interval measurements. The upper example (Figure 1a) depicts sequential sampling windows and the lower example (Figure 1b) depicts moving sampling windows.

**Sampling window**. In the interest of producing real-time metrics, dynamic metrics are often calculated over a *sampling window* of IBI measurements. The size of the sampling window, typically denoted *W*, can be defined over time or cardiac cycles.

Static metrics use a sampling window that is typically the size of an overall study period and produce a single metric result. In dynamic metrics where the size of the sampling window is less than the study period, a sequence of metrics can be produced.

**Progression**. Sampling windows of measurement can be taken by moving the sampling window forward by the size of the window (*sequential sampling windows*, Figure 1a), by a single time interval or cardiac cycle (*moving sampling windows*, Figure 1b), or any other amount. In general, we use the term *progression* for the rule which defines how the sampling window is moved.

Note in Figure 1 that **J**, the last data point, is not included in the sequential sampling windows but is included in the final moving sampling window.

Also note in Figure 1b that the initial and final *W*–1 data points of the study period participate in fewer moving windows than the more centrally-located data points. We call this characteristic of moving windows *center-bias*.

*Aggregation*. With the introduction of sampling windows, dynamic metrics can produce a sequence of results within a given study period. Those results can then be *aggregated* into a single metric for the study period. A common aggregation function (also called a *second-order function* or a *functional* in other contexts) is to take the mean of the results from each sampling window. However, other functions such as taking the standard deviation of the sampling window results might be useful in some contexts.

### Estimated Breath Cycle

This study compared a dynamic metric, called estimated breath cycle (EBC), with some established metrics of HRV. EBC is uses a sampling window whose size is based on an estimation of the length of a breath cycle of the subjects.

This study uses EBC metrics based on PRV (EBC$p$) with sequential sampling windows of 10 seconds (EBC$p^{10s}$) and 16 seconds (EBC$p^{16s}$) and moving sampling windows of 10 seconds (EBC$p^{10m}$) and 16 seconds (EBC$p^{16m}$). Mean and SD functions are used to aggregate EBC metrics within a study period.

## Method

Simultaneous measurements were taken using two independent systems on a single subject that included five periods of varying breathing rates.

EBC metrics were derived from a MindDrive™ finger sensor (Discovogue Infotronics, Modena, Italy). PPG levels were sampled at 24 Hz and processed data for HR were recorded at one-second intervals. HR readings on this system are derived from a weighted average of PRV intervals that span the 10 most recent pulsebeats.

PRV metrics other than EBC were derived from a Lightstone™ system by Wild Divine, Inc. (Las Vegas, Nevada). The Lightstone system reports instantaneous pulse rate by interpolating peaks between PPG readings taken at 30 Hz (Matt Cullen, Wild Divine, Inc., personal communication, June 28, 2012).

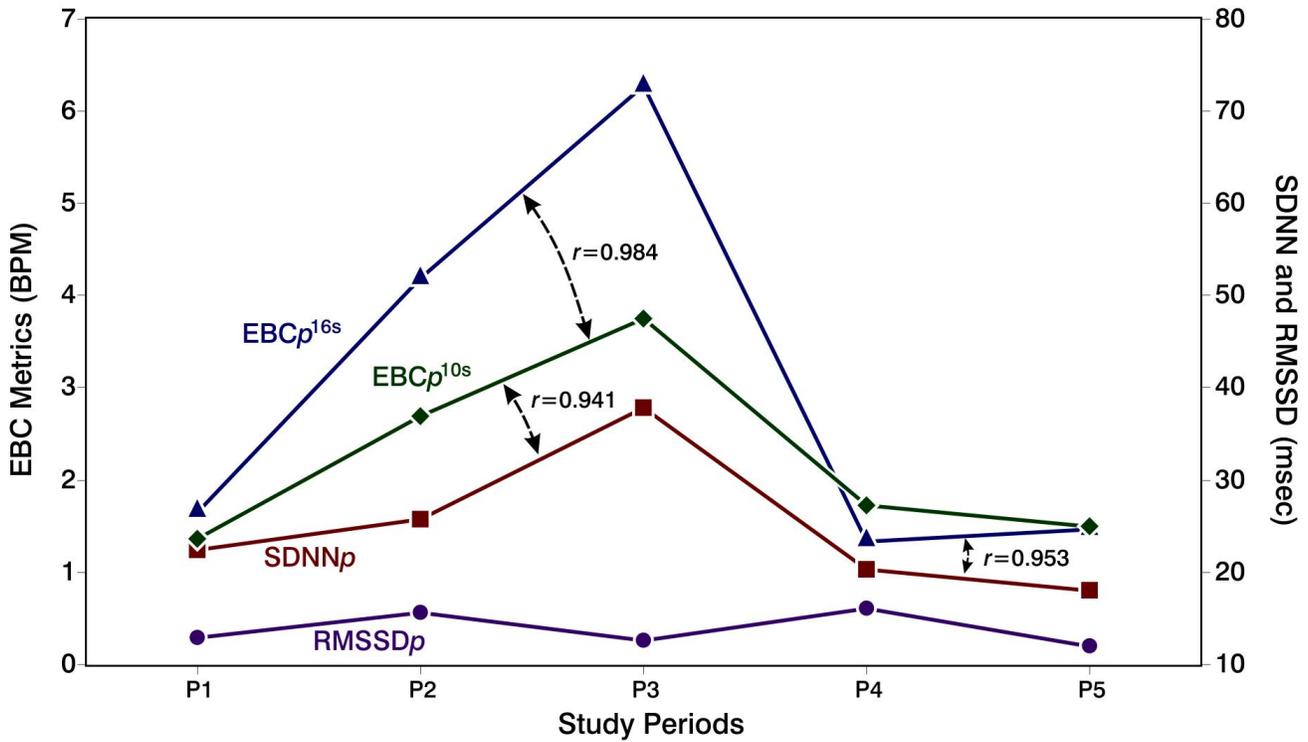

**Figure 2**. Comparison of estimated breath cycle (EBC) metrics of heart rate variability (HRV) with static HRV metrics. 10s and 16s indicate sequential windows of 10 and 16 seconds, respectively. SDNN$p$ is the standard deviation of the intervals between adjacent peaks over the study period. RMSSD$p$ is the square root of the mean of the sum of the squares of the intervals between adjacent peaks over the study period. All metrics are derived from normalized photoplethysmographic data.

*Analysis*

Artifacts were identified by visual inspection in both the MindDrive and Lightstone data streams. Time segments containing artifacts in either data stream were eliminated from the analysis.

Instantaneous pulse rate metrics from the Lightstone system were converted to IBIs, which formed the basis of all PRV metrics other than EBC.

**Results**

Figure 2 plots two of the EBC metrics against two static metrics, demonstrating a strong correlation between SDNN$p$ and EBC$p^{10s}$ ($r = .941$) and between SDNN$p$ and EBC$p^{16s}$ ($r = .953$). Numerical correlation coefficients are provided in Table 1 for all pairs of metrics.

**Discussion**

*EBC – SDNN Correlation*

The results of this pilot study demonstrated a very strong correlation between SDNN$p$ and all the EBC$p$ metrics.

The correlation between SDNN$p$ and SDNN$e$ (derived from ECG data) has been studied in three references:

**Table 1.** Pearson Correlation Coefficients between Metrics of Heart Rate Variability

| | Estimated Breath Cycle | | | |
| --- | --- | --- | --- | --- |
| | Sequential windows | | Moving windows | |
| Metric | EBC$p^{10s}$ | EBC$p^{16s}$ | EBC$p^{10m}$ | EBC$p^{16m}$ |
| SDNN$p$ | .941 | .953 | .960 | .925 |
| RMSSD$p$ | .016 | −.084 | −.027 | .042 |
| SDSD$p$ | .010 | −.001 | −.014 | .106 |
| EBC$p^{10s}$ | 1 | .984 | .997 | .990 |
| EBC$p^{16s}$ | .984 | 1 | .993 | .991 |
| EBC$p^{10m}$ | .997 | .993 | 1 | .992 |
| EBC$p^{16m}$ | .990 | .991 | .992 | 1 |

- $r = .993$ based on the average of the three conditions reported by Gil et al. (2010), Table 2;
- $r = .989$ reported by Medeiros (2010), Table 6.3; and
- $r = .99$ specifically for PRV/SDNN from a PPG finger sensor reported in Shi (2009), Table 5-4.

Since Pearson's correlation coefficients are not transitive, it is not sound practice to combine them (Castro Sotos, Vanhoof, Van Den Noortgate, & Onghena, 2008), unless a pair of coefficients, $r_a$ and $r_b$, satisfy the condition $r_a^2 + r_b^2 > 1$ (Langford, Schwertman, & Owens, 2001).

Combining the lowest SDNN$p$ – EBC$p$ correlation of $r = .925$ with lowest reported correlation of $r = .989$ for SDNN$p$ – SDNN$e$, we get a value for $r_a^2 + r_b^2$ of 1.83. Based on this, we propose that the four EBC$p$ metrics studied correlate very strongly with SDNN$e$.

### *EBC Correlation with RMSSD and SDSD*

Correlations with two additional static metric were performed:

- RMSSD$p$ is the square root of the mean of the sum of the squares of the successive differences between adjacent peaks of the PPG data over the study period.
- SDSD$p$ is the standard deviation of the successive differences between adjacent peaks of the PPG data over the study period.

Both of these metrics are estimates of the short-term / high-frequency components of HRV (Thong, Li, McNames, Aboy, & Goldstein, 2003).

Two factors of the EBC metrics of this study were expected to "smooth" the data and mask high-frequency variations: the span of 10 cardiac cycles inherent in the HR data produced by the MindDrive and the sampling window itself.

As expected, the correlation between all EBC metrics and RMSSD and between all EBC metrics and SDSD is quite low, ranging from –.084 to +.106.

### *Limitations*

This pilot study measured a small sampling of the possible dynamic HRV metrics, using a single subject, and five study periods of unequal length. We would suggest a full study in this area be carried out to further validate correlations found in this pilot study.

## Conclusions

This study tested the hypothesis that data acquisition devices based on PPG data and designed primarily for real-time HRV metrics could be useful in research studies. After characterizing the properties of the *dynamic metrics* gathered by those devices, a pilot study was carried out to compare those dynamic metrics with the metrics of HRV that are widely cited in the literature.

The results show high correlation coefficients between the EBC$p$ metrics and the established SDNN$e$. These results demonstrate the usefulness of data acquisition devices designed for real-time metrics.